\documentclass{JAC2003}
\usepackage{graphicx,amsmath}
\usepackage{booktabs}

\begin{document}
\title{COMPARISON OF COMPRESSION SCHEMES FOR CLARA}

\author{P.~H.~Williams\thanks{peter.williams@stfc.ac.uk}, J.~K.~Jones \&
  J.~W.~McKenzie\\ STFC Daresbury Laboratory, ASTeC \& Cockcroft Institute, UK}

\maketitle
\begin{abstract}
  CLARA (Compact Linear Advanced Research Accelerator) at Daresbury
  Laboratory is proposed to be the UK's national FEL test facility. The
  accelerator will be a $\sim250$~MeV electron linac capable of producing
  short, high brightness electron bunches. The machine comprises a $2.5$ cell
  RF photocathode gun, one $2$~m and three $5$~m normal conducting S-band
  ($2998$~MHz) accelerating structures and a variable magnetic compression
  chicane. CLARA will be used as a test bed for novel FEL configurations. We
  present a comparison of acceleration and compression schemes for the
  candidate machine layout.
\end{abstract}
\section{INTRODUCTION}
The design approach adopted for CLARA is to build in flexibility of operation
and layout, enabling as wide an exploration of FEL schemes as possible. For a
full overview of the aims of the project and details of FEL schemes under
consideration see~\cite{jim}.  To this end a range of possible accelerator
configurations have been considered, a selection of this work is presented
here.\par A major aim is to be able to test seeded FEL schemes. This places a
stringent requirement on the longitudinal properties of the electron bunches,
namely that the slice parameters should be nearly constant for a large
proportion of the full-width bunch length. In addition, the intention is that
CLARA has the ability to deliver high peak current bunches for SASE operation
and ultra-short pulse generation schemes, and low-emittance velocity
compressed bunches. This flexibility of delivering tailored pulse profiles
will allow a direct comparison of FEL schemes in one facility.
\section{ENERGY AT MAGNETIC COMPRESSOR}
\begin{figure}[htb]
  \centering
  \begin{tabular}{rr}
    \multicolumn{2}{c}
    {\includegraphics*[width=30mm,angle=-90]{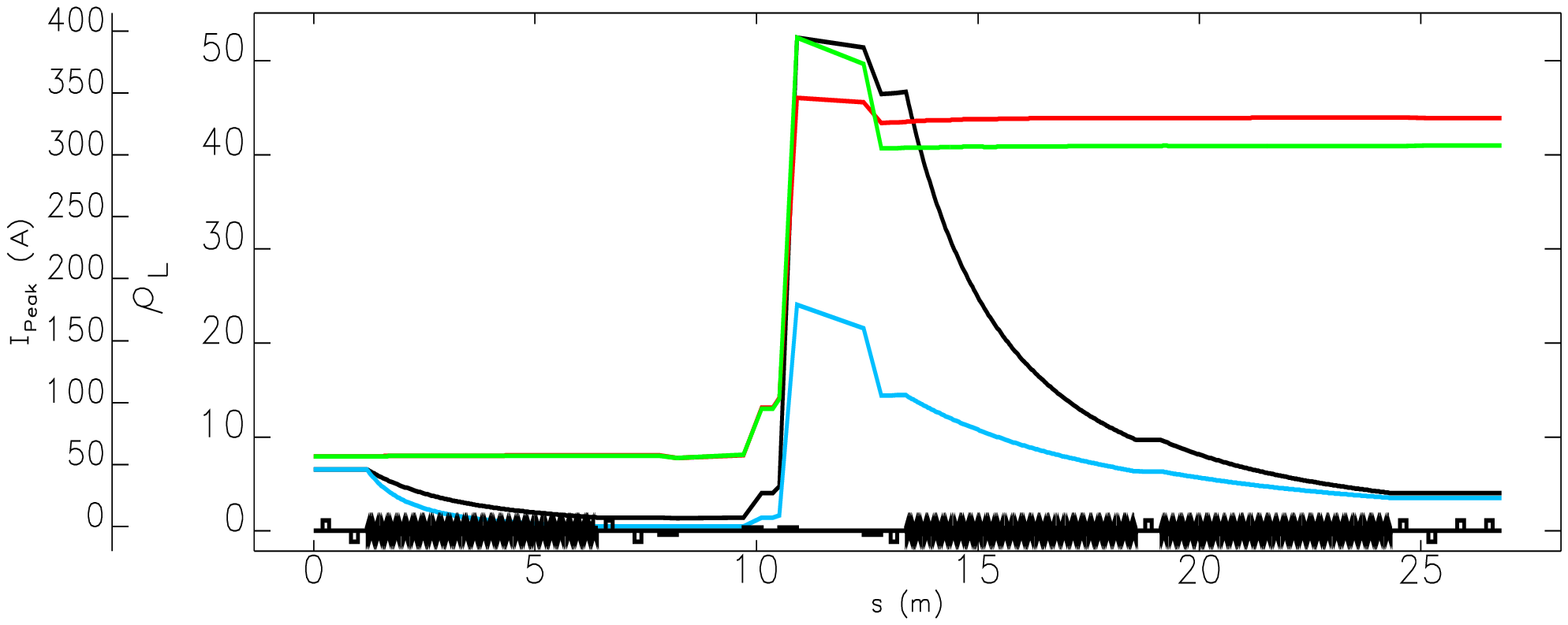}}\\
    \includegraphics*[width=30mm,angle=-90]{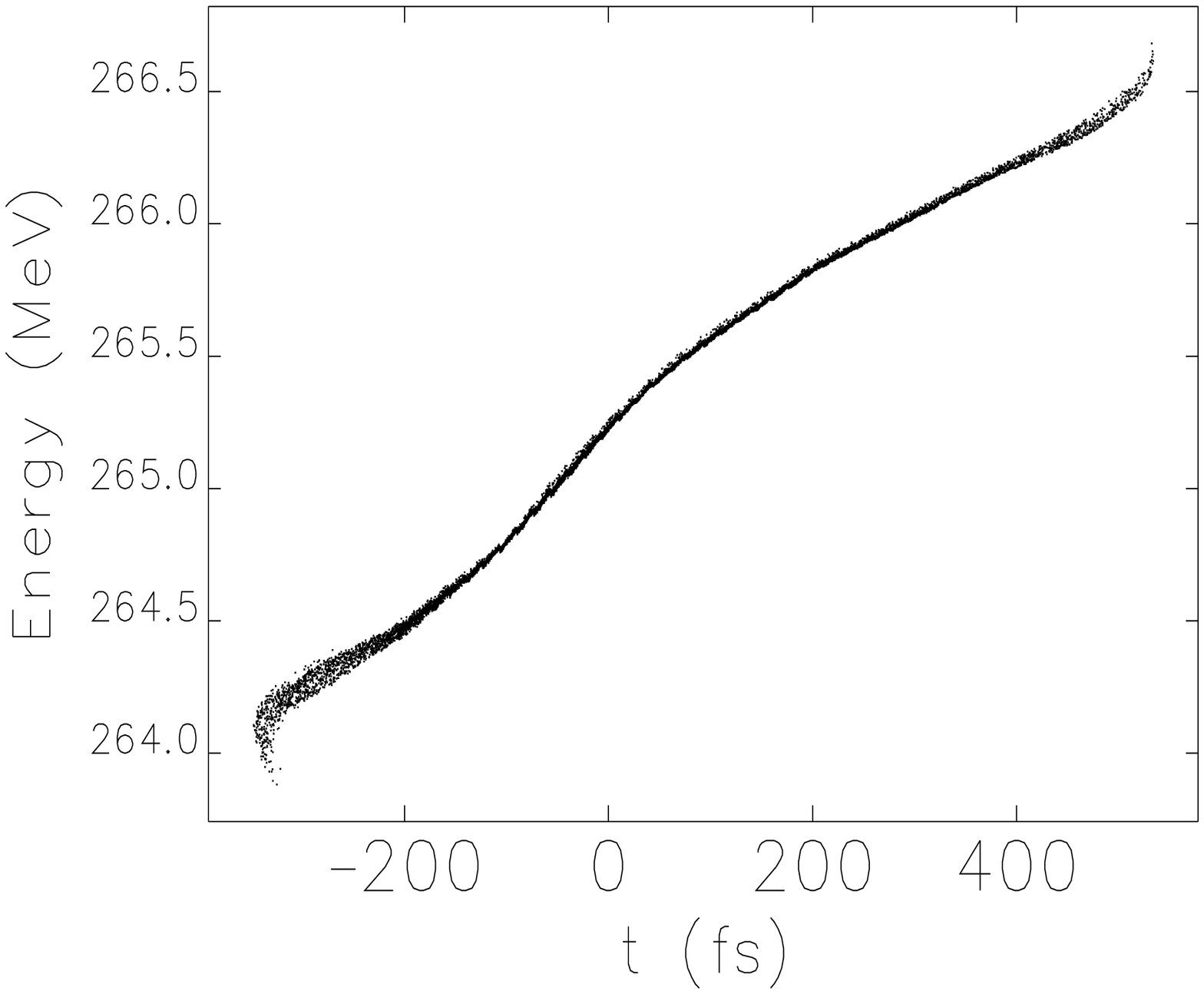}&
    \includegraphics*[width=30mm,angle=-90]{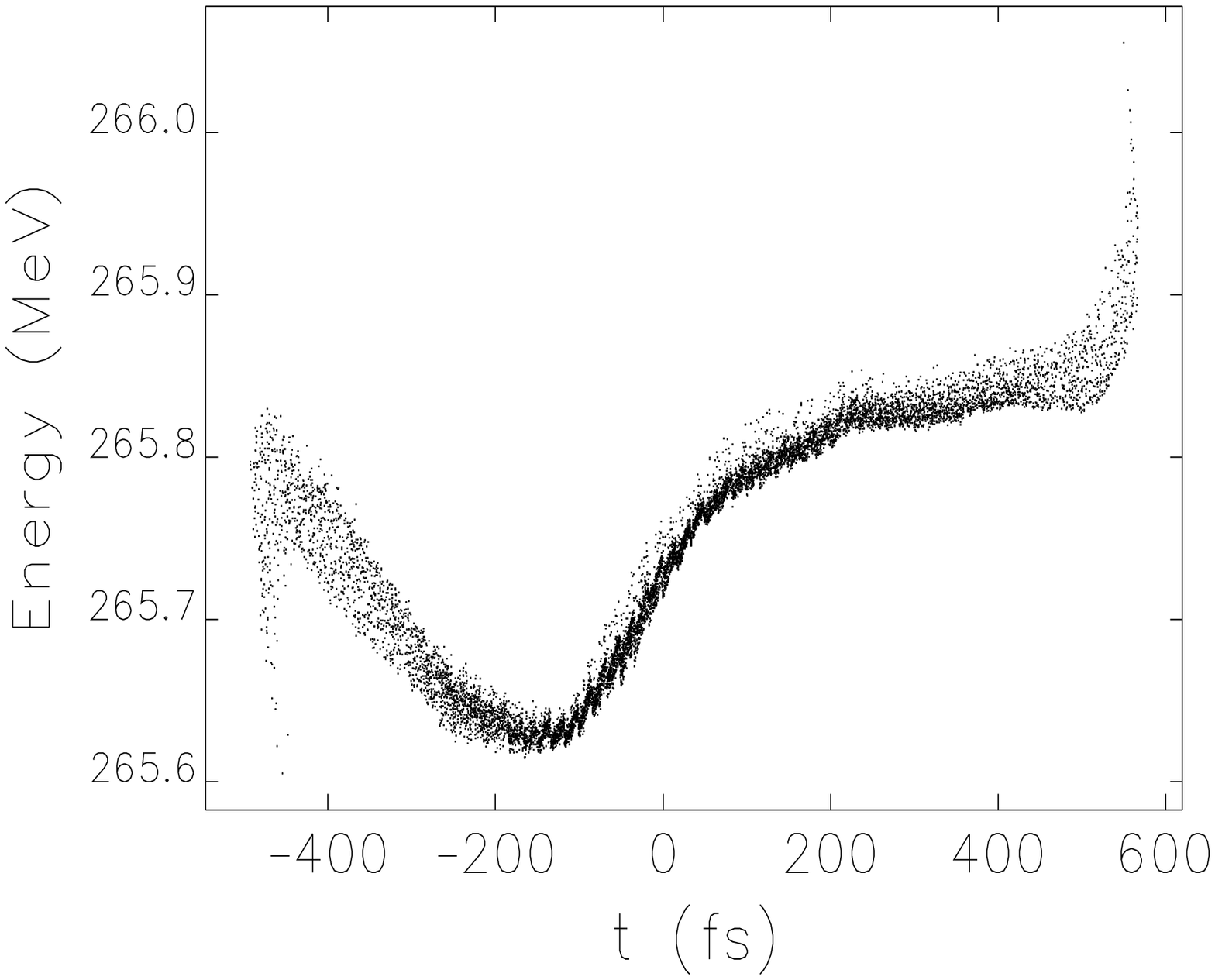}\\
  \end{tabular}
  \caption{(Upper) Laminarity (black / blue) and
    peak current (red / green) in the $10\%$ of charge slice containing the
    peak current with compression at $70$~MeV / $130$~MeV (the gun and first
    $2$~m linac module are not shown). (Lower left) Long. phase space with
    compression at $130$~MeV. (Lower right) Long. phase space with
    compression at $70$~MeV.}
  \label{fig:lamcomp}
\end{figure}

A large proportion of the FEL schemes under consideration require small
correlated energy spread at the undulators, therefore when magnetic
compression is to be used the compressor must be situated at substantially
less than full energy. This ensures that the chirp needed at compression is
able to be adiabatically damped or suppressed through running subsequent
accelerating structures beyond crest. This requirement must be balanced against
the fact that compressing at low energy exacerbates space-charge effects. To
quantify this we use the laminarity parameter
\begin{equation*}
  \label{eq:laminarity}
  \rho_L \equiv \left(\frac{I/(2I_A)}{\epsilon_{th} \gamma
  \gamma^{'}\sqrt{1/4+\Omega^2}}\right)^2,
\end{equation*}
where $I$ is the current in the slice under consideration, $I_A$ is the
Alfven current, $\epsilon_{th}$ is the thermal emittance, $\gamma^{'}\equiv
d\gamma /ds$ and $\Omega$ is a solenoidal focusing field (zero in our case).
When this parameter is greater than $1$, we should consider space-charge
effects in the bunch evolution. To inform this we select two candidate
configurations, one with magnetic compression at $70$~MeV and one at
$130$~MeV. We track a candidate $200$~pC bunch through both configurations,
setting the machine parameters attempt to produce a zero chirp bunch of peak
current $350$~A at $250$~MeV. Tracking is carried out with
ASTRA~\cite{astra,julian} to the exit of the first linac section to include
space-charge, followed by ELEGANT~\cite{elegant} taking into account the
effect of cavity wakefields, longitudinal space-charge and coherent
synchrotron radiation emittance dilution. Fig.~\ref{fig:lamcomp} shows the
resultant laminarities and final bunch longitudinal phase spaces.
We see that in both cases space-charge should be considered, this will be
achieved by re-tracking the magnetically compressed bunches in ASTRA to
elucidate any deviations due to 3-d effects as compared to the purely 1-d
effects included in ELEGANT. However compression at $130$~MeV does not allow
us to subsequently de-chirp the bunch, note that we go no further than
$30^\circ$ beyond crest in the final accelerating structures in order to
avoid large jitter effects. As we wish the facility to be flexible we select
a nominal compressor energy of $70$~MeV, but we achieve this by reducing the
gradient in the accelerating structure before the compressor. This gives us
the option of compressing at higher energy in regimes where a de-chirped
bunch is not required. With the above considerations in mind we define
engineering specifications for the CLARA variable bunch compressor as shown
in Table~\ref{tab:bcspec}. For flexibility, the compressor has a continuously
variable $R_{56}$ and is rated for maximum energy of $150$~MeV. The ability
to set a straight through path also allows investigation of purely velocity
compressed bunches.
\begin{table}[!tbp]
  \caption{Specification of variable bunch compressor.}
  \centering
  \begin{tabular}{lrl}
    \toprule
    \textbf{} & \textbf{Value} & \textbf{Unit} \\
    \midrule
    Energy at compressor & 70 - 150 & MeV\\
    Min. : Max. bend angle & 0 : 200  & mrad\\
    Bend magnetic length & 200 & mm\\
    Max. bend field & 0.5 & T\\
    Min. : Max. transverse offset & 0 : 300 & mm\\
    Z-distance DIP-01/04 - DIP-02/03 & 1500 & mm\\
    Z-distance DIP-02 - DIP-03 & 1000 & mm\\
    Max. bellows extension & 260 & mm\\
    Min. : Max. $R_{56}$ & 0 : -72 & mm\\
    Max. $\sigma_x$ from $\delta_E$ ($\pm 3\sigma$) & 0 : 10 & mm\\
    Max. $\sigma_x$ from $\beta_x$ ($\pm 3\sigma$) & 1.5 & mm\\
    \bottomrule
  \end{tabular}
  \label{tab:bcspec}
\end{table}
\section{BUNCH FOR SEEDING SCHEMES}
A seeded FEL scheme requires constant bunch parameters over a large
proportion of the bunch. This reduces the sensitivity to timing jitter
between the seed laser and electron bunch. Specifically, we require a
constant peak current of $350$~A over $300$~fs of the bunch, with zero chirp,
constant emittances and zero transverse offset. In order to achieve this we
must cancel the curvature that originates from RF acceleration. It is
possible to do this purely magnetically, although typically it is done with
higher harmonic RF. As harmonic RF entails additional expense we compare two
schemes, a bunch compressor with non-linear elements inserted, and a fourth
harmonic X-band cavity.
\subsection{LINEARISATION VIA NONLINEAR CHICANE}
The lower right plot of Fig.~\ref{fig:lamcomp} shows residual curvature in
the longitudinal phase space. This can be flattened by changing the sign of
the natural $T_{566}$ term in the bunch compressor chicane. To achieve this
sextupoles were added to the chicane. The number, positions and strengths of
these were parameters of an optimisation. We impose the constraints that the
$T_{566}<20$~cm, the derivative of dispersion with respect to energy and it's
derivative with respect to $s$ should be zero on exit of the chicane, the
projected emittances should not exceed $1$~mm mrad and the sextupoles
$k_{2}<2000$~m$^{-2}$.  This constraint set was chosen by trial and error.
\begin{figure}[htbp]
  \centering
  \includegraphics*[width=85mm]{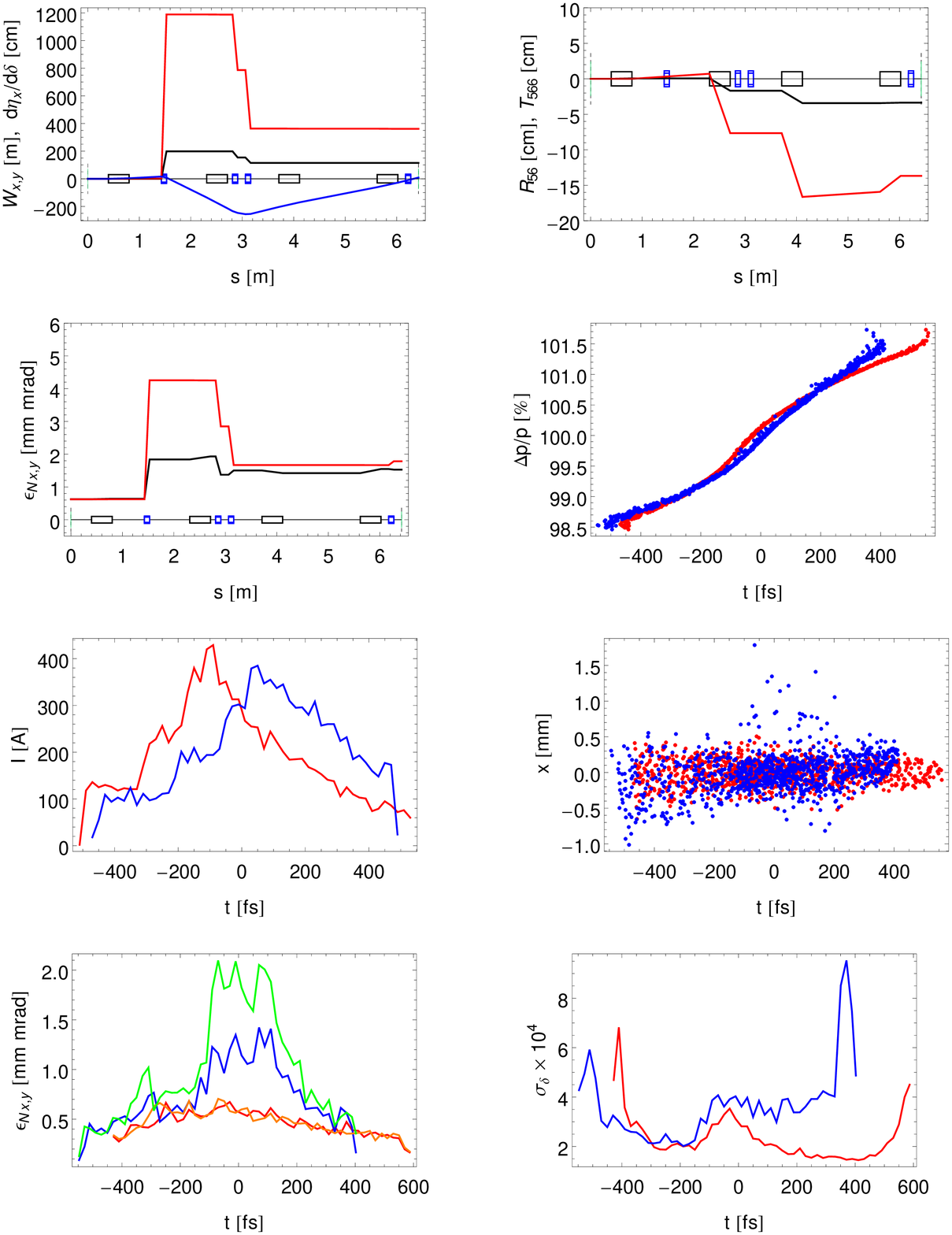}
  \caption{Results for an example optimised nonlinear
    bunch compressor. (1) Chromatic amplitude functions (black, red) \&
    chromatic derivative of dispersion (blue). (2) $R_{56}$ (black) \&
    $T_{566}$ (red).  (3) $\varepsilon_{N(x,y)}$ less dispersive
    contributions (black, red). (4) Longitudinal phase space (blue -
    optimised, red - without sextupoles). (5) Current profile ($20$~fs
    slices, blue - optimised, red - without sextupoles). (6) $x-t$ phase
    space (blue - optimised, red - without sextupoles). (7) Normalised slice
    emittances ($20$~fs slices): horizontal (blue - optimised, red - without
    sextupoles) and vertical (green - optimised, orange - without
    sextupoles). (8) Slice energy spread ($20$~fs slices) (blue - optimised,
    red - without sextupoles)}
  \label{fig:nonlinbcres}
\end{figure} 
Figure~\ref{fig:nonlinbcres} shows the optimisation results. Flattening the
longitudinal curvature is relatively straightforward however the chromatic
properties are easy to spoil, resulting in increased projected and slice
emittance. Up to six sextupoles were tried with similar results. These
nonlinear compressors have also been studied under energy jitter and the
bunch parameters found to vary substantially.
\subsection{LINEARISATION VIA HARMONIC RF}
\begin{figure}[htbp]
  \centering
  \includegraphics*[width=85mm]{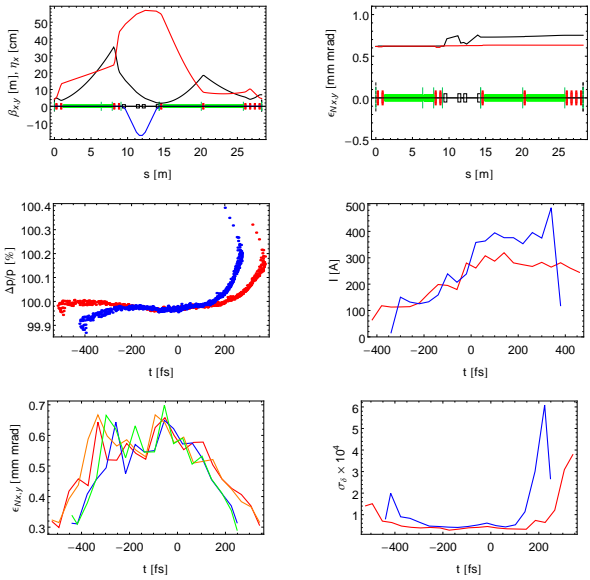}
  \caption{Two candidate optimisations linearising with harmonic cavity. 
    (1) Optics: $\beta_{x,y}$ (black, red) \& $\eta_x$. (2)
    $\varepsilon_{N(x,y)}$ less dispersive contributions (black, red).  (3)
    Longitudinal phase space (blue - optimised for $200$~fs flat top, red -
    optimised for $300$~fs flat top). (4) Current profile ($40$~fs slices,
    optimised for $200$~fs flat top, red - optimised for $300$~fs flat top).
    (5) Normalised slice emittances ($40$~fs slices): horizontal / vertical
    (blue / red - optimised for $200$~fs flat top, green / orange - optimised
    for $300$~fs flat top) (6) Slice energy spread ($20$~fs slices) optimised
    for $200$~fs flat top, red - optimised for $300$~fs flat top.}
  \label{fig:4hcres}
\end{figure}

We insert a fourth harmonic $0.7$~m structure immediately prior to the
magnetic compressor. An optimisation~\cite{lj} was then performed with
variables being the harmonic voltage and phase, the off crest phase of the
preceding linac and the angle of the compressor dipoles. Results for two
candidate tunings are shown in Fig.~\ref{fig:4hcres}. The peak voltage on the
linearising cavity is $7$~MV/m. It can be seen that the additional
complication of a harmonic cavity is justified by ability to predictably
tailor longitudinal phase space.

\section{VELOCITY BUNCHING}
An alternative to magnetic compression is to use velocity bunching in the low
energy section of the accelerator. The first $2$~m linac section is set to
the zero crossing to impart a time-velocity chirp along the bunch. The bunch
then compresses in the following drift space. The second linac section is
positioned at the waist of the bunch length evolution after $3$~m of drift to
rapidly accelerate the beam and capture the short bunch length.  Solenoids
are required around the bunching section to control the transverse beam size
and prevent emittance degradation. ASTRA was used to track until the end of
the second linac module followed by ELEGANT. The quadrupoles between the
first and second linac sections are switched off in order to keep the beam
axially symmetric, and the bunch compressor set to zero angle. An
evolutionary algorithm was used to optimise the beamline for both bunch
length and transverse emittance.  We present two tunings with $100$~pC bunch
charge.
\begin{figure}[htb]
  \centering
  \includegraphics*[width=60mm,angle=-90]{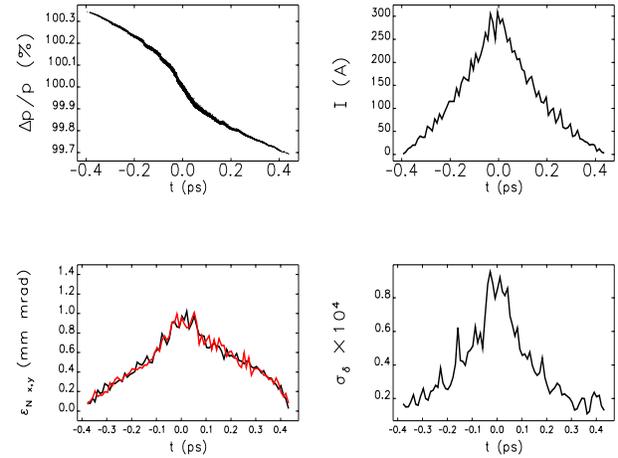}
  \caption{Velocity bunched beam tuned for low emittance.}
  \label{fig:vblowemit}
\end{figure}
\begin{figure}[htb]
  \centering
  \includegraphics*[width=60mm,angle=-90]{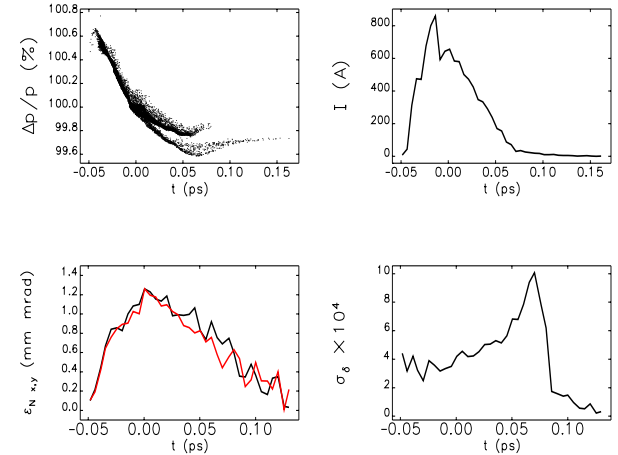}
  \caption{Velocity bunched beam tuned for peak current.}
  \label{fig:vbpeak}
\end{figure}
Fig.~\ref{fig:vblowemit} shows a bunch with similar peak current and current
profile to the non-linearised magnetically compressed bunch of
Fig.~\ref{fig:nonlinbcres}-5. This is achieved at half the total bunch
charge, with lower slice energy spread, but higher slice emittance.  In
Fig.~\ref{fig:vblowemit} we show that a similar peak current to the
non-linearised magnetically compressed bunch of Fig.~\ref{fig:nonlinbcres}-5
is easily achieved with smaller slice energy spread but higher slice
emittance. Fig.~\ref{fig:vbpeak} shows a beam tuned for peak current at the
exit of the second linac module. The peak current then degrades along the
accelerator. This bunch has the capabilities to provide single-spike SASE FEL
operation.
\section{CONCLUSIONS}
This initial study has established an accelerator layout for CLARA that is
inherently flexible in the pulse profiles it is capable of producing. We have
shown this by simulating bunches suitable for seeded and SASE FEL
operation. Further work will entail jitter tolerance analysis of the
presented configurations.

\end{document}